\documentclass{sig-alternate}
\usepackage{setspace}

\usepackage{graphicx}
\usepackage{algorithmicx}
\usepackage[ruled,boxed,algonl,vlined]{algorithm2e}
\usepackage{subfigure}

\usepackage{graphics}
\begin{document}
%

\title{Thresholding of Semantic Similarity Networks using a Spectral Graph Based Technique}
%
%
%
%
%

\numberofauthors{4} 
%
\author{
%
%
\alignauthor Pietro Hiram Guzzi\\
       \affaddr{Department of Surgical and Medical Sciences}\\
       \affaddr{University of Catanzaro}\\
       \affaddr{Italy}\\
       \email{hguzzi@unicz.it}
\alignauthor Simone Truglia\\
       \affaddr{Department of Surgical and Medical Sciences}\\
       \affaddr{University of Catanzaro}\\
       \affaddr{Italy}\\
       \email{trgsimo@gmail.com}
\and
\alignauthor Pierangelo Veltri\\
       \affaddr{Department of Surgical and Medical Sciences}\\
       \affaddr{University of Catanzaro}\\
       \affaddr{Italy}\\
       \email{veltri@unicz.it}
\alignauthor Mario Cannataro\\
       \affaddr{Department of Surgical and Medical Sciences}\\
       \affaddr{University of Catanzaro}\\
       \affaddr{Italy}\\
       \email{cannataro@unicz.it}
       }


\maketitle
\begin{abstract}

Semantic similarity measures (SSMs) refer to a set of algorithms used to quantify the similarity of two or more terms belonging to the same ontology. Ontology terms may be associated to concepts, for instance in computational biology gene and proteins are associated with terms of  biological ontologies. Thus, SSMs may be used to quantify the similarity of genes and proteins starting from the comparison of the associated annotations. SSMs have been recently used to compare genes and proteins even on a system level scale. More recently some works have focused on  the building and analysis of Semantic Similarity Networks (SSNs) i.e. weighted networks in which nodes represents genes or proteins while weighted edges represent the semantic similarity score among them. SSNs are quasi-complete networks, thus their analysis presents different challenges that should be addressed. For instance, the need for the introduction of reliable thresholds  for the elimination of meaningless edges arises. Nevertheless, the use of global thresholding methods may produce the elimination of meaningful nodes, while the use of local thresholds may introduce biases. For these aims, we introduce a novel technique, based on spectral graph considerations and on a mixed global-local focus. The effectiveness of our technique is demonstrated by using markov clustering for the extraction of biological modules. We applied clustering  to simplified networks  demonstrating a considerable improvements with respect to the original ones.

\end{abstract}




\section{Introduction}


The  accumulation of raw experimental data about genes and proteins  has been accompanied by the accumulation of functional information, i.e. knowledge about function.  The assembly, organization and analysis of this data has given a considerable impulse to research \cite{cannataro2013data}.
Usually biological knowledge is encoded by using annotation terms, i.e. terms describing for instance function or localization of genes and proteins. Such annotations are often organized into ontologies, that offer a formal framework to organize in a formal way biological knowledge \cite{Guzzi2012}. For instance, Gene Ontology (GO) \cite{Harris:GeneOntology:NAR2004} provides a set of annotations (namely GO Terms) of biological aspects, structured into three main taxonomies: Molecular function (MF), Biological Process (BP), and Cellular Component (CC). Annotations are often stored in publicly available databases, for instance a main resource for GO annotations is the Gene Ontology Annotation (GOA) database \cite{GOA:citeulike:461337}.

A set of algorithms, referred to as Semantic Similarity measures (SSMs), enabled the comparison of set of terms belonging to the same ontology. SSMs take in input two or more ontology terms and produce as output a value representing their similarity. This enabled the possibility to use such formal instruments for the comparison and analysis of proteins and genes \cite{Guzzi2012}.

Consequently, many works have focused on: (i) the definition of ad-hoc semantic measures tailored to the characteristics of Gene Ontology ; (ii) the definition of measures of comparison among genes and proteins; (iii) the introduction of methodologies for the systematic analysis of metabolic networks; (iv) building of \textit{semantic similarity networks}, i.e. edge-weighted graph whose nodes are genes or proteins, and edges represent semantic similarities among them \cite{Pesquita2009}.

A semantic similarity network of proteins (SSN) is an edge-weighted graph $G_{ssu}$=$(V$,$E)$, where $V$ is the set of proteins, and $E$ is the set of edges, each edge has an associated weight that represent the semantic similarity among related pairs of nodes.

These networks are constructed by computing some similarity value between genes or proteins. Nevertheless, such networks are usually quasi complete networks, so the use of them as framework of analysis has many problems.

Thus the definition of a threshold on the edge weight to retain only the meaningful relationships is a crucial step. An high threshold may result on the loss of many significant relationship while a low threshold may introduce a lot of noise-

In other kind of networks many methods have been defined: for instance the use of an arbitrary global threshold \cite{freemanthreshold}, or the use only of a fraction of highest relationship \cite{alathreshhold}, or statistical based methods \cite{Rito15092010}.  Nevertheless, internal characteristics of SSMs (as investigated in \cite{guzzi2012cimento}) do not suggest the use of global thresholds. In fact small regions of relatively low similarities may be due to the characteristics of measures while proteins or genes have high similarity.  Thus the use of local threshold may constitute an efficient way, i.e retaining only top k-edges for each node \cite{lee2004,moryama2003,voy2006}. Although this consideration, this choice may be influenced by the presence of local noise and in general may cause the presence of biases in different regions.

Starting from these considerations, we developed a novel hybrid method that merges together both local and global considerations. This method is based on spectral graph theory and it is based on two main considerations.

We apply a local threshold for each node, i.e we retain only edges whose weight is higher than the average of all its adjacent. The choice of the threshold is made by considering a global consideration: the emergence   of nearly-disconnected components by looking at the laplacian of the graph and its eigenvalues \cite{ding2001,ngspectralclustering}. In particular we build a novel graph in which edge weights are 0,5 and 1. The weight 0,5 is associated to edges that are retained considering only one adjacent node, while the weight 1 is associated to edges that are retained.

The choice of this simplification has a biological counterpart on the structure of biological networks. It has been proved in many works that these biological networks tend to have a modular structure in which hubs proteins (i.e. relevant proteins) have many connections  \cite{Ma01112012,Bertolazzi2013274,Zhu01052007}. Moreover, many works proved the existence of community structures, i.e. small dense regions with few link to other regions \cite{Su15122010}. These considerations have usually inspired many algorithms for extracting biological relevant modules by analyzing biological networks \cite{10.1109/TKDE.2012.225}.

From these consideration arises the main hypothesis of this paper: the simplification of quasi complete SSN by removing non relevant edges to evidence the formation of a structure of networks characterized by relatively-small dense networks loosely coupled with other ones.

After the application of the proposed simplification,  we analyze resulting networks by applying a common algorithms used to mine graphs. We show that thresholded networks have in general more performances and that the best ones are reached with nearly-disconnected ones.

\section{Problem Statement}

We here introduce main concepts used for the formulation of the main problem of this article.

 \subsection{Spectral Graph Analysis}

Spectral graph theory \cite{chung1994} refers to the study of the properties of a graph by looking at the properties of the eigenvalues and eigenvectors of matrices associated to the graph. In particular we here focus on the Laplacian matrix of a graph that is defined as follows \cite{cvetkovic2010towards,Bolla19941}.

Given an edge-weighted graph $G$ with $n$ nodes,  we may define the weighted adjacency matrix $A$ as the $nxn$ matrix in which the element $a_{i,j}$ is defined as follows.

\begin{equation}\label{weighted}
    a_{i,j}=\left\{
              \begin{array}{ll}
                w_{i,j}  & \hbox{if i,j are connected;} \\
                0, & \hbox{if i,j are not connected }
              \end{array}
            \right.
\end{equation}

For these graphs the notion of degree may be easily extended in this way. For each vertex $v_i$ the degree is defined as the sum of the weights of all the adjacent edges $vol_{v_i}=\Sigma_{j} w_{i,j}$.  Then  we may define the Degree Matrix $D$ as follows:

\begin{equation}\label{weighted}
    d_{i,j}=\left\{
              \begin{array}{ll}
                vol_{v_i} , & \hbox{if i=j;} \\
                0, & \hbox{elsewhere}
              \end{array}
            \right.
\end{equation}

Finally, the Laplacian Matrix $L$ is defined as $L=D-A$. Similarly in literature other slightly definitions of Laplacian (e.g. Signless Laplacian, Normalized Laplacian \cite{merris1994laplacian}) have been proposed.


Beside the other properties that are related to the characteristic polynomial of laplacian, we here focus on the smallest nonzero eigenvalue, often referred to as Fiedler vector \cite{citeulike:9080319}. It has been shown that the number of connected components is related to the algebraic multiplicity of the smallest eigenvalues in case of both un-weighted and weighted graphs. Starting from this consideration, Ding et al. \cite{ding2001} observed that also nearly-disconnected components may also identified by analyzing the eigenvector associated to the Fiedler vector.

For this study we analysed the spectrum of the graph obtained after the simplification under the hypothesis that a graph with nearly disconnected component may represent a suitable choice. If the graph is connected we will build a novel graph. If the graph has-nearly disconnected component we end the process and we mine the resulting subgraph for the identification of biological relevant modules.

\subsection{Semantic Similarity Measures}

A semantic similarity measure ($SSMs$) is a formal instrument to quantify the similarity of two or more terms of the same ontology. Measures comparing only two terms are often referred to as pairwise semantic measures, while measures that compare two sets of term yielding a global similarity among sets are referred to as groupwise measures.

Since proteins and genes are associated to a set of terms coming from Gene Ontology,  $SSMs$ are often extended to proteins and genes. Similarity of proteins is then translated in the determination of similarity of set of associated terms \cite{Couto:SEMANTICSIMILARITIBIOMEDONTO:PLOS2009,citeulike:7730206}.
Many similarity measures have been proposed (see for instance \cite{Guzzi2012} for a complete review) that may be categorized according to different strategies used for evaluating similarity. We here do not discuss deeply $SSMs$ for lack of space.

\section{The Proposed Approach}

We here introduce a method for threshold selection on weighted graph based on the spectrum of the associated laplacian matrix. The process is straightforward. The pruning algorithm examines each node in the input graph. For each node it stores all the weights of the adjacent edges. Then it determines a local threshold $k=\mu + \alpha * sd$, where $\mu$ is the average of weights, $sd$ is the standard deviation and $\alpha$ is a variable threshold that is fixed globally. In this way we realize an hybrid approach since the threshold $k$ has a global component $\alpha$ and a local one given by the average and standard deviation of the weights of the adjacent.

 If the weight of an edge is greater than $k$ considering the adjacent of both its nodes, then it will be inserted into the novel graph with unitary weight. Otherwise, hen if the weight of an edge is greater than $k$ considering only one of its adjacent nodes, then it will be inserted into the novel graph with weight 0,5. At the end of this process, the Laplacian of the spectrum of the graph is analyzed as described in Ding et al \cite{ding2001}. If the graph presents nearly disconnected components, then the process stops, alternatively a novel graph with a more stringent threshold $k$ is generated.

\subsection{Building Semantic Similarity Networks}

Following algorithm explains the building of the semantic similarity network $G_{ssu}$ by iteratively calculating semantic similarity among each pair of proteins. For each step two proteins are chosen and the semantic similarity among them is calculated. Then nodes are added to the graph and an edge is inserted when the semantic similarity is greater than 0.

\begin{algorithm}{Building Semantic Similarity Networks}
 \SetAlgoLined
 \KwData{Protein Dataset P, Semantic Similairity Measure SS }
 \KwResult{Semantic Similarity Network $G_{ssu}$=$V_{ssu},E_{ssu}$}
 initialization\;
 \ForAll{ $p_i$ in P }{
  read $p_i$\;
  add $p_i$ in $V_{ssu}$ \;
 \ForAll{$p_j$ in P, $j \neq i$ }
  {Let $\sigma$=SS($p_i$,$p_j$) \;
  \If{alpha is greater than 0}{
   add the weighted edge ($p_i$,$p_j$,$\sigma$) to $E_{ssu}$\;
   }}
 }
 \caption{Building Semantic Similarity Networks}
\end{algorithm}

%

\subsection{Pruning Semantic Similarity Networks}

This section explains the pruning of semantic similarity network through an example. To better clarify the process, we use an auxiliary graph $G_{pr}$ that is the final process of pruning. The graph is built in an incremental fashion by considering all the nodes of $G_{ssu}$.  The process is straightforward. The pruning algorithm examines each node $  \in G_{ssu}$. For each node it stores all the weights of the adjacent edges. Then it determine a local threshold (for instance the average of the weights or the median value as exposed after). At the end of this step, the node $i$ and all the adjacent ones are inserted in to $G_{pr}$ (only if they are not yet present).

Then each edge adjacent to $i$ with weight greater with the determined local threshold is inserted into $G_{pr}$. If the considered edge is not present in $G_{pr}$, the edge will have weight 0.5, otherwise the weight of the edge is set to 1. We used in this work two simple thresholds, the average and the median of all the weights. Finally all the nodes with 0 degree are deleted from $G_{pr}$.

The rationale of this process is that edges that are \textit{relevant} considering the neighborhood of both nodes will compare in the pruned graph with unitary weight while edges that are \textit{relevant} considering one node will compare with 0.5 weight. In this way we think that we may reduce the noise.

\begin{figure}[ht]
  \centering
 \includegraphics[width=5in]{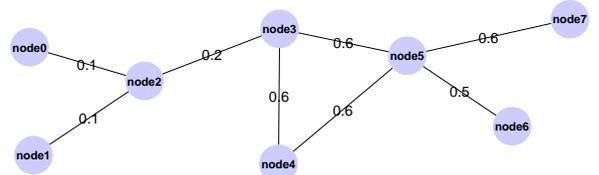}
  \caption{Weighted Semantic Similarity Network.}
  \label{fig:esempio1}
\end{figure}

For instance, let us consider the network depicted in Figure \ref{fig:esempio1} and let us suppose that threshold is represented by the average. Without loss of generality we suppose k=0 in this example. Let $AVG(node_i)$   be the average of the weights of nodes adjacent to $node_i$ that is used as threshold.

 \begin{itemize}
 \item The algorithm initially explores $node_0$, since it has degree 1, it is discarded from the analysis.
 \item Then it explores $node_1$ that is discarded similarly to $node_0$.
 \item When $node_2$ is considered, the algorithm adds into $G_{pr}$ $node_0, node_1,node_2,$ and $node_4$ and the edge $(node_2,node_4)$ with weight 0.5 - (the average of the weights of the neighbours of $node_2$ is equal to 0,13 and other two edges have a lower weight). Figure \ref{fig:esempio1visitanode2} depicts the produced graph at this step.

\begin{figure}[ht]
  \centering
 \includegraphics[width=2.5in]{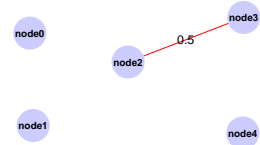}
  \caption{The output of the algorithm at Step 2}
  \label{fig:esempio1visitanode2}
\end{figure}

 \item $node_3$ is reached by the visiting. Then $node_4$, and $node_5$ are inserted into $G_{pr}$. The $AVG(node_3)$ is equal to 0,46, so only edges ($node_3, node_5$) and $(node_3,node_5)$ are inserted into $G_{pr}$ with weight 0.5. Figure \ref{fig:esempio1visitanode3} depicts $G_{pr}$ after this step.

\begin{figure}[ht]
  \centering
 \includegraphics[width=2.5in]{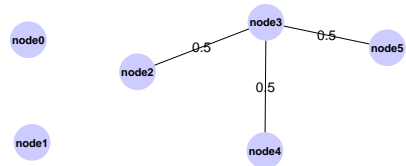}
  \caption{Output  after the visit of node3.}
  \label{fig:esempio1visitanode3}
\end{figure}

 \item $node_4$ is reached. Since all the adjacent nodes have been inserted into $G_{pr}$, no nodes are added into this step. The $AVG(node_4)$ is equal to 0,6, so all the edges must be inserted. In particular edge $(node_4,node_3)$ is yet present, so its weight is updated to 1.0. Diversely, $(node_4,node_5)$ is inserted with weight equal to 0.5. Figure \ref{fig:esempio1visitanode4} depicts $G_{pr}$ after this step.

\begin{figure}[h]
  \centering
 \includegraphics[width=2.5in]{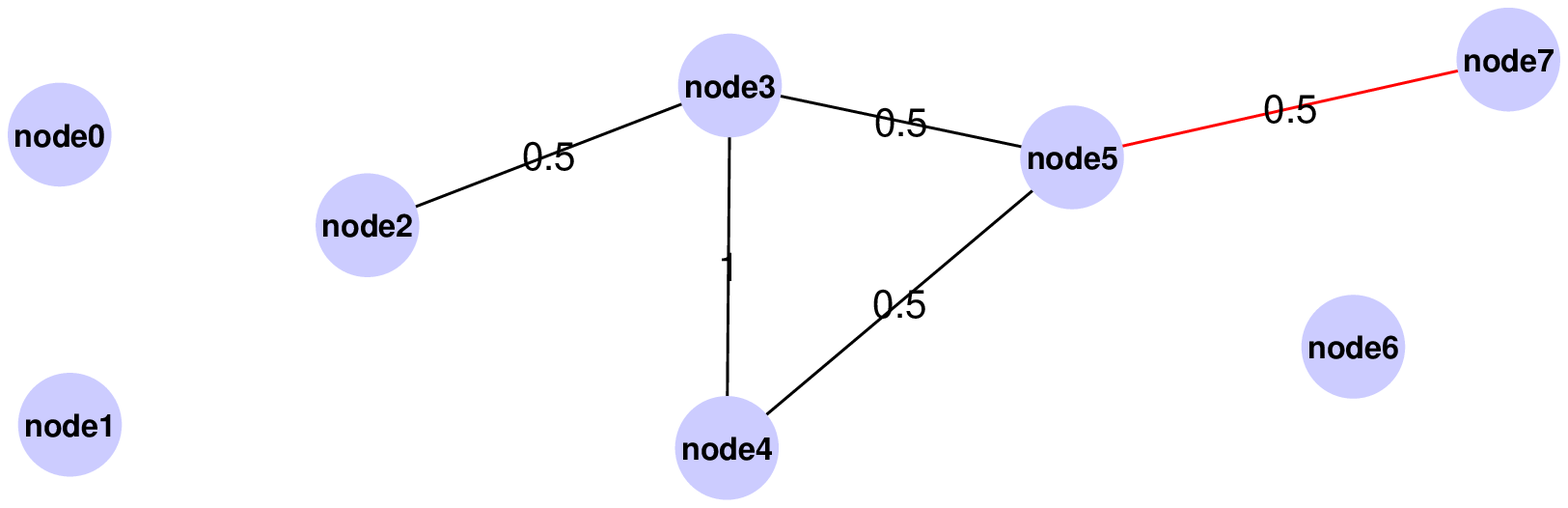}
  \caption{Output after the visit of node4.}
  \label{fig:esempio1visitanode4}
\end{figure}

 \item $node_5$ is reached. $node_7$ and $node_8$  are inserted into    $G_{pr}$. The  $AVG(node_5)$ is 0,575.  Consequently the weight of $(node_5,node_3)$ $(node_5,node_34$  and in $G_{pr}$  is updated to 1,  $(node_6,node_7)$ is inserted into $G_{pr}$.Figure \ref{fig:esempio1visitanode5} depicts $G_{pr}$ after this step.

\begin{figure}[h]
  \centering
 \includegraphics[width=2.5in]{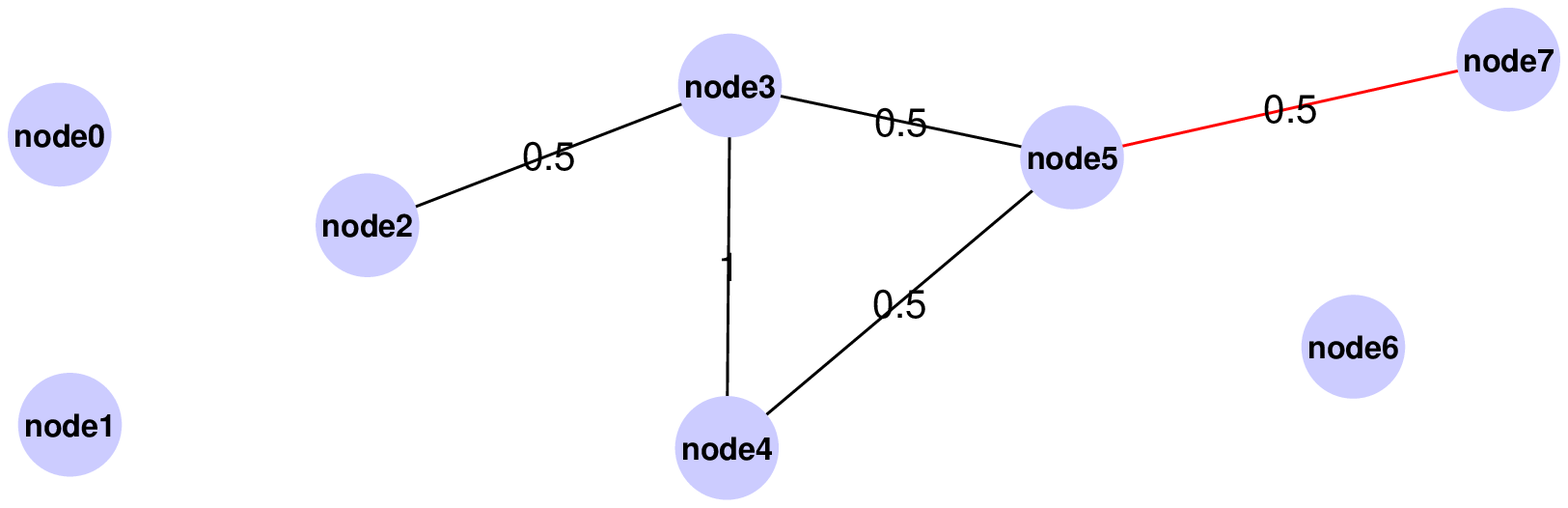}
  \caption{Output after the visit of node5.}
  \label{fig:esempio1visitanode5}
\end{figure}

  \item $node_7$ and $node_8$ are visited but discarded since they have degree equal to 1.

\item Finally all the nodes with zero degree are eliminated from $G_{pr}$, producing the resulting graph depicted in Figure \ref{fig:esempio1visitanode5}.

\begin{figure}[h]
  \centering
 \includegraphics[width=2.5in]{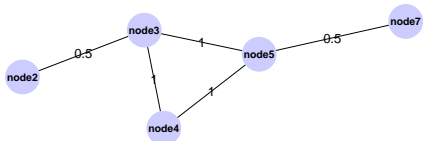}
  \caption{Final pruned graph}
  \label{fig:esempio1visitanode5}
\end{figure}
\end{itemize}


The generation of pruned graph is repeated until the graph has nearly disconnected components. This may be evident by analyzing the spectrum of the associated laplacian for value of threshold.
%
%
%
\begin{verbatim}
Pruning Semantic Similarity Network
 Input SSn Raw Semantic Simlarity Network,
 K  Threshold of Simplification
 Output: SSp Simplified Semantic Similarity Network
  While SSp has not nearly-disconnected component
 		$SSp$ = Simplify(SSn,k)
   Increment k
Return: SSp
 \end{verbatim}

%

%
%
%

\subsection{Analysis of Semantic Similarity Networks}

As introduced, in a Semantic Similarity Networks, nodes represent proteins or genes, and edges represent the value of similarity among them. Starting from a dataset of genes or proteins, a SSN may be built in an iterative way, and once built, algorithms from graph theory may be used to extract topological properties that encode biological knowledge.

As starting point, the global topology of an semantic similarity network, i.e. the study of the clustering coefficient or of the diameter, can reveal main properties of the network and the correspondence with respect to a theoretical model.

In addition to analysis of global properties, the study of recurring local topological features and the extraction of relevant modules, i.e. cliques, has found an increasing interest. For the purposes of this work, we focus on the extraction of dense subgraphs  under the hypothesis that they could encode protein complexes.

SS measures are able to quantify the functional similarity of pairs of proteins/genes, comparing the GO terms that annotate them. Thus, there are no constraints on the minimum set size \cite{Guzzi2012}.

Since proteins within the same pathway are involved in the same biological process, they are likely to have high semantic similarity. In a similar way, protein belonging to the same complex are likely to have similar biological roles, and therefore they should have high semantic similarity.

The rationale of this study is to demonstrate the ability of semantic similarity networks to represent in a similar way to protein interaction networks. Main difference is represented by the fact that semantic similarity networks may encode more knowledge that is hidden in protein interaction networks.


There exist currently main approaches of analysis of protein interaction networks that span a broad range, from the analysis of a single network by clustering to the comparison of two or more networks trough graph alignment approaches \cite{DBLP:journals/fgcs/CannataroGV10,DBLP:journals/csur/CannataroGV10}. In this work we consider the use of Markov Clustering Algorithm (MCL) as mining strategy. MCL has been proved to be a good predictor of functional modules when applied to protein interaction networks.

\section{Case Study}

%
%

In order to show the effectiveness of this strategy we propose the following assessment:

\begin{itemize}
\item we downloaded  three dataset of proteins (the CYC2008 dataset \footnote{wodaklab.org/cyc2008/}, the MIPS catalog \cite{16381906}, and the Annotated Yeast High-Throughput Complexes \footnote{wodaklab.org/cyc2008/} );
\item  we calculated different semantic similarities among them using FastSemSim tool \footnote{fastsemsim.sourceforge.net} (we considered 10 semantic similarity measures from those available in FastSemSim ( Czekanowsky-Dice , Dice, G-Sesame, Jaccard, Kin, NTO,  SimGic, SimICND, SimIC, SimUI, TO
\cite{Cannataro:2010:ACM.1824796} ) and two ontologies Biological Process (BP) and Molecular Function (MF). Consequently we generated 20 SSN for each input dataset.
\item we applied the pruning of the semantic network with varying threshold  causing the presence of nearly disconnected components and the presence of disconnected components;
\item we extracted modules on the raw and simplified networks at various threshold showing the improvements of our strategy showing the improvement in terms of functional enrichment of modules (i.e. the quantification of biological meaning of modules).
\end{itemize}

 As final step we compare our simplification with other global strategies demonstrating the effectiveness of the local simplification.

\subsection{Results}

For each generated network we used the markov clustering algorithm (MCL) to extract modules. The effectiveness of the use of MCL for detecting modules in networks has been demonstrated in many works (see for instance \cite{Cannataro:2010:ACM.1824796}). We here assess how MCL is able to discover \textit{functionally coherent} modules in different semantic similarity network and how this process is positively influenced by the simplification. In particular we show how the process of simplification improves the overall results and how best results are obtained when networks presents nearly disconnected components.


We evaluated the obtained results in terms of \textit{functional coherence} of extracted modules. We define \textit{functional coherence } $FC$ of a module $M$ as the average of semantic similarity values of all the pair of nodes (i,j) composing a module. 
\begin{displaymath}\sum_{i,j}^{} \frac{SSM(i,j)}{N}\end{displaymath}, where N is the number of the proteins of the module.


%
%

Starting from this definition, we may obtain a single value for all the modules extracted in an execution of MCL by averaging these values. We consider this average value as a representative for the thresholded network.

 \begin{figure*}[ht]
  \centering
 \includegraphics[width=5in]{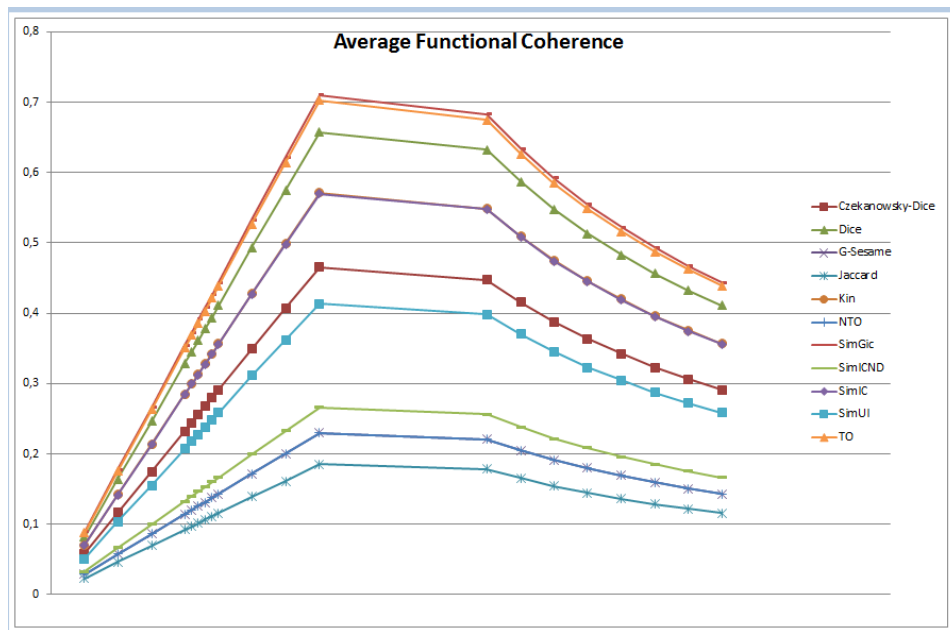}
  \caption{Comparison of Average FC at different Threshold Levels on CYC2008 Dataset}
  \label{fig:comparison}
\end{figure*}

 \begin{figure*}[ht]
  \centering
 \includegraphics[width=5in]{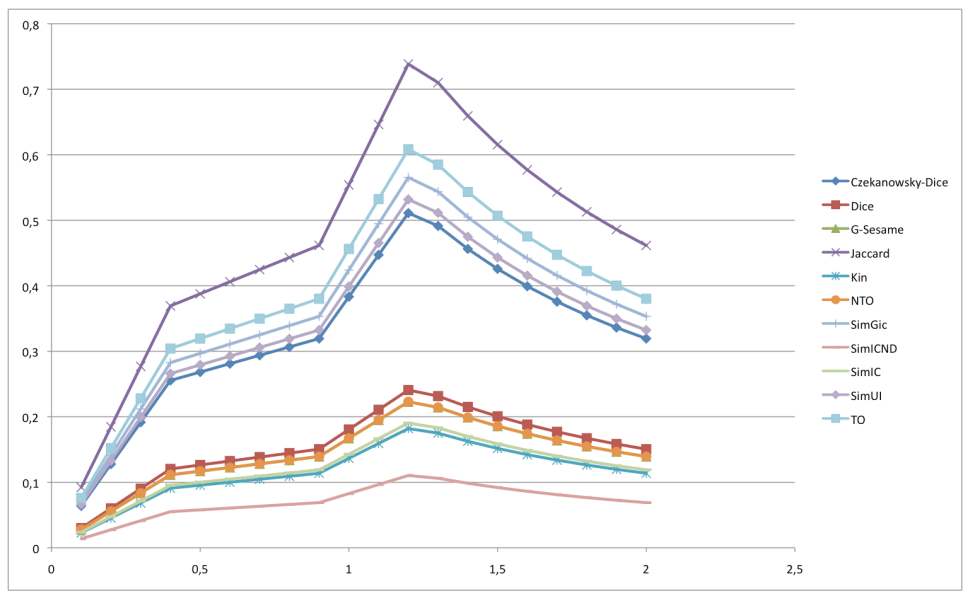}
  \caption{Comparison of Average FC at different Threshold Levels on MIPS }
  \label{fig:comparison}
\end{figure*}

 \begin{figure*}[ht]
  \centering
 \includegraphics[width=5in]{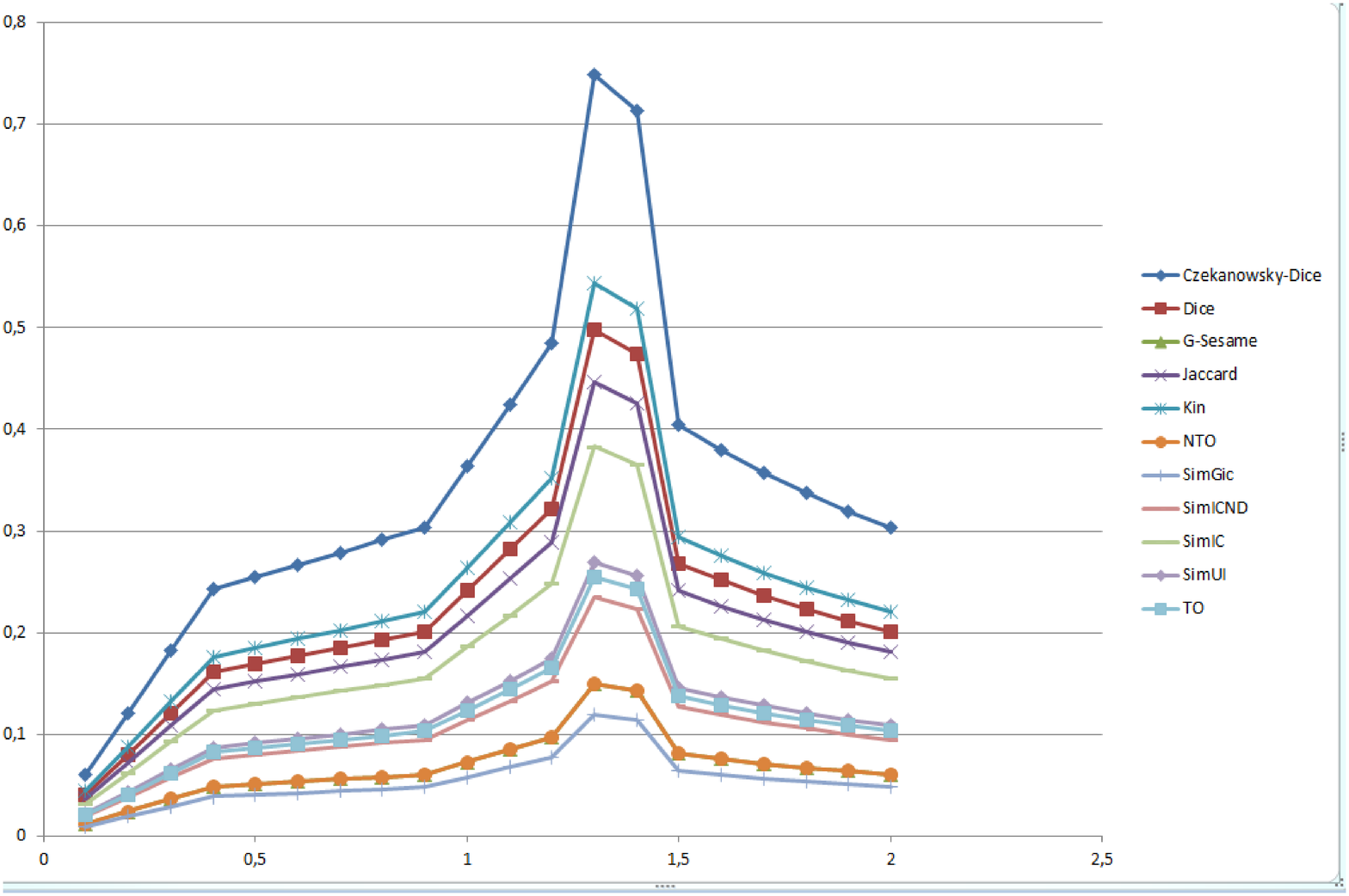}
  \caption{Comparison of Average FC at different Threshold Levels on Annotated High Throughput Complexes Dataset}
  \label{fig:comparison}
\end{figure*}

\section{Conclusion}

Results showed that raw semantic similarity networks contains lot of noise, thus are unsuitable for the analysis. Consequently we proposed a local simplification of networks. Result confirm that mining of simplified networks is a suitable way for extract biologically meaningful knowledge.

\bibliographystyle{amsplain}
\bibliography{ref}

\end{document}